\documentclass[letterpaper, 11pt]{article}

\usepackage{amsmath, amssymb, amsfonts, amsthm}
\usepackage{bm}
\usepackage{mathtools}
\usepackage{mathrsfs}
\usepackage{geometry}
\usepackage{setspace}
\usepackage[utf8x]{inputenc}
\usepackage{algpseudocode,algorithmicx,algorithm}
\usepackage[T1]{fontenc}
\usepackage{lmodern}
\usepackage{graphicx}
\usepackage{enumerate}
\usepackage[natural, dvipsnames, pdftex]{xcolor}
\usepackage{tikz}
\usepackage{verbatim}
\usepackage{hyperref}
\usepackage{cancel}
\usepackage[all]{xy}
\usepackage{mdframed}
\usepackage[capitalize]{cleveref}
\usepackage{todonotes}
\usepackage{algpseudocode}
\usepackage{comment}

\usepackage[numbers]{natbib}

\usepackage{enumitem}

\usepackage{indentfirst}
\theoremstyle{plain}
\newtheorem{thm}{Theorem}[section]
\newtheorem{lemma}[thm]{Lemma}

\newtheorem{cor}[thm]{Corollary}

\theoremstyle{definition}
\newtheorem{defn}[thm]{Definition}

\theoremstyle{remark}
\newtheorem{rmk}[thm]{Remark}

\newcommand{\F}{\mathbb{F}}

\newcommand{\cC}{\mathcal{C}}

\newcommand{\Vol}{\mathrm{Vol}}

\begin{document}

\title{A $\Theta(n)$-Multiplicative Improvement of the Gilbert-Varshamov Bound for Linear and Quantum Codes
\thanks{
C. Yuan is with School of Computer Science, Shanghai Jiao Tong University. (Email: \href{chen_yuan@sjtu.edu.cn}{chen\_yuan@sjtu.edu.cn}) R. Zhu is with School of Computer Science, Shanghai Jiao Tong University. (Email: \href{sjtuzrq7777@sjtu.edu.cn}{sjtuzrq7777@sjtu.edu.cn})
}}

\author{Chen Yuan, Ruiqi Zhu}
\date{}
\maketitle

\begin{abstract}
The \emph{Gilbert-Varshamov} (GV) bound is a fundamental benchmark for the existence of error-correcting codes. For non-linear $q$-ary codes of code length $n$, its finite length form is known to admit a multiplicative improvement of order $\Theta(n)$.
In contrast, for $q$-ary linear codes of arbitrary rates, the best previously known general improvement was only of order $\Theta(\sqrt n)$. This leaves open whether the algebraic constraint of linearity inherently prevents a linear-order refinement. We show that it does not. For every fixed prime power $q$ and relative distance $0<\delta<1-1/q$, we obtain a $\Theta(n)$-multiplicative improvement of the GV bound for $q$-ary linear codes of arbitrary rate, matching the corresponding result for nonlinear codes. Our proof is based on a simple change of viewpoint. Rather than requiring the initial random linear construction to contain no low-weight vectors, we allow $O(n)$ low-weight points, show that they are independent in the underlying projective space, and then eliminate all of them simultaneously at the cost of a single dimension. The key probabilistic ingredient is an exponential bound on nontrivial linear dependencies among low-weight vectors.

We further show that this approach can be extended to quantum codes despite the symplectic self-orthogonality constraint. We obtain a $\Theta(n)$-multiplicative improvement of the quantum GV bound for every fixed $0<\delta<1-1/q^2$ satisfying $H_{q^2}(\delta)<1/2$. Thus, the same $\Theta(n)$ finite length improvement is applicable to both general linear structure and symplectic self-orthogonality.

\end{abstract}

\section{Introduction}
The \emph{Gilbert-Varshamov} (GV) bound is one of the fundamental existential benchmarks in coding theory. It gives a general trade-off between the rate and the minimum distance of a code. Let $A_q(n,d)$ denote the maximum size of a $q$-ary code of length $n$ and minimum distance $d$, and let $\mathrm{Vol}_q(n,d)=\sum_{i=0}^{d}\binom{n}{i}(q-1)^i$ be the volume of a $q$-ary Hamming ball of radius $d$. The classical GV bound asserts that $A_q(n,d)\ge\frac{q^n}{\mathrm{Vol}_q(n,d-1)}.$ The bound originates from the work of Gilbert and Varshamov~\cite{gilbert1952comparison,varshamov1957estimate} and has since become one of the basic reference points in coding theory; see, e.g.,~\cite{macwilliams1977theory,pless1998handbook}. Variants and extensions of the GV argument have been studied in a wide range of settings~\cite{gu1993generalized,marcus1992improved,tolhuizen1997generalized} and the bound also continues to serve as a benchmark in the study of structured, efficiently encodable, locally testable, and locally correctable codes~\cite{brehm2026linear,gopi2018locally,sendrier2004linear,shum2001low,kasami1974gilbert}.

Improving the GV bound is one of the central problems in coding theory. For square alphabet sizes $q\geq 49$, algebraic-geometric constructions of Tsfasman, Vl\u{a}du\c{t}, and Zink can go beyond the GV bound~\cite{tsfasman1982modular}; see also~\cite{xing2003nonlinear}. 
However, the GV bound is still the asymptotically best known lower bound for codes over small alphabets. In this work, we study \emph{finite length multiplicative refinements} of the GV bound which can be dated back to \cite{jiang2004asymptotic}. Jiang and Vardy obtained the first $\Theta(n)$-multiplicative improvement for non-linear binary codes.  Vu and Wu~\cite{vu2005improving} subsequently extended this result to $q$-ary codes with the same improvement.

The problem is more challenging for the GV bound of the linear codes. A $\Theta(n)$-multiplicative improvement was previously known only for binary linear code of rate-$1/2$~\cite{gaborit2008asymptotic}. Gaborit and Z\'{e}mor~\cite{gaborit2008asymptotic} constructed a family of binary random double-circulant codes with the same linear-multiplicative improvement as non-linear code. However, their result is restricted to rate $1/2$, due to its double-circulant construction.  Their argument relies heavily on the particular cyclic structure of the code ensemble, it is not clear how to extend it to $q$-ary linear codes or to binary codes of arbitrary rates. Therefore, the method of~\cite{gaborit2008asymptotic} fails to achieve $\Theta(n)$-multiplicative improvement for $q$-ary linear codes of arbitrary rate.

For $q$-ary linear codes of arbitrary rate, Hao, Huang, Livshyts, and Tikhomirov~\cite{hao2022distribution} obtained a $\Theta(\sqrt n)$-multiplicative improvement based on the method of moments. Yuan and Zhu~\cite{yuan2026improvement} obtained an improvement of the same order via an elementary probabilistic approach. Thus, the best known multiplicative improvements for general linear codes remain $\Theta(\sqrt n)$. 

Since the GV bound characterize the existence of codes, we can also obtain the GV bound for the quantum codes. Quantum error-correcting codes play a central role in protecting quantum information against noise and decoherence and are indispensable for fault-tolerant quantum computation~\cite{shor1995scheme,knill1998resilient,kitaev2003fault,terhal2015quantum,leverrier2022quantum,panteleev2022asymptotically} and reliable quantum communication~\cite{schumacher1996quantum,barnum1998information,lloyd1997capacity}. A standard approach to constructing stabilizer quantum codes is through classical codes satisfying an appropriate self-orthogonality condition~\cite{steane1996error,aly2007quantum,bierbrauer2000quantum,grassl1999quantum,ketkar2006nonbinary}. In particular, a symplectic self-orthogonal subspace $\cC\subseteq \cC^{\perp_S}\subseteq\mathbb F_q^{2n}$ gives rise to a $q$-ary stabilizer code whose minimum distance is controlled by the symplectic distance of $\cC^{\perp_S}$~\cite{ashikhmin2001nonbinary}. Quantum GV bounds and their finite length refinements have been studied in~\cite{2000quantum,feng2004finite,jin2011quantum}. Concatenated quantum codes attaining the asymptotic quantum GV bound were constructed in~\cite{ouyang2014concatenated}, while explicit quantum codes over $q\in\{2,3,4,5\}$ exceeding the finite GV bound~\cite{feng2004finite} were constructed in~\cite{guan2022construction}. Asymptotic improvements beyond the quantum GV bound for some relative distances were also obtained from algebraic-geometry codes for $q=p^{2r}$ with an odd prime $p$ in~\cite{niehage2007nonbinary}, and for square prime powers $q\ge 19^2$ in~\cite{feng2006asymptotic}. Recently, Yuan and Zhu~\cite{yuan2026improvement} showed that a $\Theta(\sqrt n)$-multiplicative improvement can also be obtained for quantum codes with symplectic self-orthogonality. The self-orthogonality constraint makes their probabilistic argument more challenging. 


In summary, $\Theta(\sqrt n)$-multiplicative improvement is still the state-of-the-art result for $q$-ary linear codes of arbitrary rate and quantum codes over small-sized alphabet. This raises a natural question. Is it possible to obtain $\Theta(n)$-multiplicative improvement for both $q$-ary linear codes of arbitrary rate and quantum codes? In this paper, we answer this question affirmatively. We show that $\Theta(n)$-multiplicative improvements is attainable for  $q$-ary linear codes of arbitrary rate and for quantum codes under symplectic self-orthogonality. Thus, at the level of these finite length refinements, neither linearity nor symplectic self-orthogonality causes a loss in the best known order of improvement compared with non-linear codes.

\subsection{Our Contributions and Techniques}

Our first result gives a $\Theta(n)$-multiplicative improvement of the Gilbert-Varshamov bound for $q$-ary linear codes. A common approach is to construct a linear code one dimension at a time, avoiding low-weight codewords at each step. In contrast, we start with $k+1$ message dimensions and allow low-weight codewords to appear. Using the exponential bound of Yuan and Zhu~\cite{yuan2026improvement} and counting minimally dependent sets, we show that the corresponding low-weight directions are linearly independent with positive probability. We can then eliminate all these directions with a single linear constraint, obtaining a $k$-dimensional code with minimum distance at least $d$.

Let $G\in\mathbb F_q^{(k+1)\times n}$ be a uniformly random matrix. Note that a vector $\mathbf{c}$ and any nonzero scalar multiple $\lambda\mathbf{c}$ have the same weight. Thus, we define projective space $\mathbb{P}^{k}(\F_q)\coloneqq (\F_q^{k+1}\setminus\{\mathbf{0}\})\slash\F_q^*$ and check if $S(G):=\{[\mathbf{x}]\in\mathbb P^k(\mathbb F_q): \mathbf{x}G\in B_q(n,d-1)\}$ is empty. If $S(G)$ is empty, then we are done as $G$ is a generator matrix of some linear code with minimum distance at least $d$. Otherwise, we divide it into two cases, $S(G)$ is projectively independent or projectively dependent. $S(G)=\{[\mathbf{x}_1],\ldots,[\mathbf{x}_t]\}$ is said to be projectively independent if $\mathbf{x}_1,\ldots,\mathbf{x}_t$ are $\F_q$-linearly independent and otherwise projectively dependent.

If $S(G)$ is projectively dependent, then it contains a minimally projectively dependent subset $\{[\mathbf{x}_1],\ldots,[\mathbf{x}_r],[\mathbf{y}]\}$.
Then our probabilistic argument will bound the probability that the random vectors $\mathbf{x}_1G,\ldots,\mathbf{x}_rG$ lie in $B_q(n,d-1)$ and meanwhile their linear combination $\mathbf{y}G$ also lie in $B_q(n,d-1)$. Conditioning on that $\mathbf{x}_1G,\ldots,\mathbf{x}_rG$ are uniformly at random in $B_q(n,d-1)$, Corollary~\ref{cor:coefficients} shows that $\mathbf{y}G$ also lies in the ball with probability at most $e^{-\gamma n}$. We take a union bound over all such minimally projectively dependent subsets to show that $S(G)$ is projectively dependent with low probability. Then, we move to the case that $S(G)=\{[\mathbf{x}_1],\ldots,[\mathbf{x}_t]\}$ is projectively independent. In this case, we note that $\mathbf{x}_1,\ldots,\mathbf{x}_t$ are $\F_q$-linearly independent and $t\leq k+1$. Define a linear map $\varphi: \F_q^{k+1}\rightarrow \F_q$ such that $\varphi(\mathbf{x}_i)=1$ for $i\in [t]$. The kernel space $V$ has dimension $k$ and become our message space. Let $\cC$ be the linear code with generator matrix $G$.  Using $V$ as the message space yields a subcode $\cC'\subseteq \cC$ of dimension $k$ which does not contain $\mathbf{x}_1,\ldots,\mathbf{x}_t$. In this way, we show the existence of $k$-dimensional linear code $\cC'$ with length $n$ and minimum distance $d$. This proves the existence of the desired $k$-dimensional linear code.

Our second result is a $\Theta(n)$-multiplicative improvement of the quantum Gilbert-Varshamov bound. To prove this result, we start with a uniformly random $(k-1)$-dimensional symplectic self-orthogonal subspace $\cC\subseteq\mathbb F_q^{2n}$. For $\mathbf{v},\mathbf{w}\in\cC^{\perp_S}\setminus\cC$, we say that $\mathbf{v}$ and $\mathbf{w}$ are $\cC$-equivalent if $\mathbf{v}-\lambda\mathbf{w}\in\cC$ for some $\lambda\in\mathbb F_q^*$, and denote the equivalence class of $\mathbf{v}$ by $[\mathbf{v}]_{\cC}$. We call $[\mathbf{v}]_{\cC}$ bad if it contains a nonzero vector in $B_q^S(2n,d-1)$. Let $\mathcal B(\cC)$ be the collection of all bad $\cC$-equivalence classes. $\{[\mathbf{v}_1]_{\cC},\ldots,[\mathbf{v}_t]_{\cC}\}$ is said to be a $\cC$-independent set if $\lambda_1\mathbf{v}_1+\cdots+\lambda_t\mathbf{v}_t\in\cC$ implies $\lambda_1=\cdots=\lambda_t=0$ and otherwise $\cC$-dependent set.

It is easy to show that $\cC\cap B_q^S(2n,d-1)=\{\mathbf{0}\}$ via a probabilistic argument. However, the minimum distance of a quantum code depends on its dual code $\cC^{\perp_S}$, so this condition alone is not enough. Rather than requiring $\mathcal B(\cC)$ to be empty, we show that $\mathcal B(\cC)$ is $\cC$-independent with positive probability. Unlike the Hamming linear code case, a nonzero linear combination of low-weight vectors may lie in $\mathcal C$, even when these vectors are linearly independent. If $\mathcal B(\cC)$ is $\cC$-dependent, then it contains a minimally $\cC$-dependent set $\{[\mathbf{v}_1]_{\cC},\ldots,[\mathbf{v}_t]_{\cC},[\mathbf{y}]_{\cC}\}$, where $\mathbf{v}_1,\ldots,\mathbf{v}_t,\mathbf{y}$ are low weight vectors. There are two cases. If these vectors are linearly dependent, we handle this case as in the linear code argument. Otherwise, there exists a nonzero $\mathbf{c}\in\cC$ such that $\mathbf{y}-\mathbf{c}\in\mathrm{span}\{\mathbf{v}_1,\ldots,\mathbf{v}_t\}$. Since the first $t$ classes are $\cC$-independent, we have $\mathrm{span}\{\mathbf{v}_1,\ldots,\mathbf{v}_t,\mathbf{y}\}\cap\cC =\langle\mathbf{c}\rangle$. This allows us to apply Lemma~\ref{lem:quotient-dependence}. Together with the bound for the first case and the bound on the number of bad classes, we show that with positive probability $\mathcal B(\cC)$ is $\cC$-independent and $\cC$ contains no nonzero low weight vector. Lemma~\ref{lem:one-step-elimination} then extends $\cC$ by one dimension to a $k$-dimensional symplectic self-orthogonal subspace $\cC'\supseteq\cC$ whose dual $\cC'^{\perp_S}$ contains no nonzero low-weight vector. This proves the desired minimum distance.

\subsection{Organization}

In Section~\ref{sec:2}, we introduce the basic notation and preliminary results for classical linear codes, the Hamming metric, and symplectic self-orthogonal codes. In Section~\ref{sec:3}, we prove the $\Theta(n)$-multiplicative improvement of the Gilbert-Varshamov bound for $q$-ary linear codes. In Section~\ref{sec:4}, we extend the argument to symplectic self-orthogonal codes and obtain $\Theta(n)$-multiplicative improvement of the quantum Gilbert-Varshamov bound.

\section{Preliminaries}\label{sec:2}

Throughout the paper, let $q$ be a prime power and let $\F_q$ denote the finite field with $q$ elements. For any finite nonempty set $S$, let $\mathrm{Unif}(S)$ denote the uniform distribution over $S$. We first recall some standard notations for classical linear codes and the Hamming metric, and then introduce the symplectic strucutre for quantum codes.

\subsection{Linear codes}
A $q$-ary linear code $\cC$ of length $n$ and dimension $k$ is a $k$-dimensional subspace of $\F_q^n$. More generally, for a not necessarily linear code of length $n$ over $\F_q$ with $M$ codewords, its rate is defined as $R=\frac{\log_q M}{n}.$ In particular, the rate of an $[n,k]_q$-linear code is $R=k/n$.

For two vectors $\mathbf{x}=(x_1,\ldots,x_n),\,\mathbf{y}=(y_1,\ldots,y_n)\in\F_q^n,$ their Hamming distance is $d(\mathbf{x},\mathbf{y}):=\bigl|{i\in[n]:x_i\neq y_i}\bigr|.$ The minimum distance of a code $\cC$ is $d(\cC):=\min_{\substack{\mathbf{x},\mathbf{y}\in\cC\ \mathbf{x}\neq\mathbf{y}}}d(\mathbf{x},\mathbf{y}).$ The Hamming weight of $\mathbf{x}$ is $\mathrm{wt}(\mathbf{x}):=d(\mathbf{x},\mathbf{0}).$ For a linear code, it holds that $d(\cC)=\min_{\mathbf{0}\neq\mathbf{x}\in\cC}\mathrm{wt}(\mathbf{x}).$ A linear code of length $n$, dimension $k$, and minimum distance $d$ is called an $[n,k,d]_q$-linear code. A code with minimum distance $d$ can correct up to $\lfloor(d-1)/2\rfloor$ errors. The relative minimum distance of an $[n,k,d]_q$ code is $d/n$. 

The $q$-ary entropy function $H_q(x):=x\log_q(q-1)-x\log_q x-(1-x)\log_q(1-x)$  is strictly increasing on $0<x<1-1/q$.

\begin{defn}[Hamming Ball]
For an integer $d\geq 0$, define the $q$-ary Hamming ball $$B_q(n,d):=\{\mathbf{x}\in\F_q^n:\mathrm{wt}(\mathbf{x})\leq d\}.$$ 
Let 
$\Vol_q(n,d):=|B_q(n,d)|=\sum_{i=0}^{d}\binom{n}{i}(q-1)^i$ be its volume and for every fixed
$0<\delta<1-1/q$, it holds that 
$$q^{H_q(\delta)n-o(n)} \leq \Vol_q(n,\lfloor\delta n\rfloor)\leq q^{H_q(\delta)n}.$$
\end{defn}

For integer $1\leq k<n$, we choose a uniformly random matrix $G\in \F_q^{(k+1)\times n}.$ Define 
$\mathbb{P}^{k}(\F_q)\coloneqq (\F_q^{k+1}\setminus \mathbf{0})\slash \F_q^*$ to be the projective space of dimension $k$. For any nonzero $\mathbf{x}\in \F_q^{k+1}$, let $[\mathbf{x}]:=\{\lambda \mathbf{x}:\lambda\in \F_q^*\}$ to be an element in $\mathbb{P}^{k}(\F_q)$. Define
\begin{equation}\label{eqn:S_G}
    S(G):=\{[\mathbf{x}]\in\mathbb{P}^{k}(\F_q) : \mathbf{x}G\in B_q(n,d-1)\}.
\end{equation}
For every fixed $[\mathbf{x}]\in\mathbb{P}^{k}(\F_q)$, it is clear that the random vector $\mathbf{x}G$ is uniformly distributed in $\F_q^n$. One can show that 
$$\Pr[\mathbf{x}G\in B_q(n,d-1)]=\frac{\Vol_q(n,d-1)}{q^n}, \quad \mathbb E\bigl[|S(G)|\bigr]=\frac{(q^{k+1}-1)\Vol_q(n,d-1)}{q^n(q-1)}$$

\subsection{Quantum Codes}
 The symplectic structure used in the construction of quantum codes. Given $\mathbf{a}=(a_1,\ldots,a_n),\,\mathbf{b}=(b_1,\ldots,b_n),\,\mathbf{a}'=(a_1',\ldots,a_n'),\,\mathbf{b}'=(b_1',\ldots,b_n')$, the symplectic inner product on $\F_q^{2n}$ is  $$\left\langle(\mathbf{a}\mid\mathbf{b}),(\mathbf{a}'\mid\mathbf{b}')\right\rangle_S:=\langle\mathbf{a},\mathbf{b}'\rangle_E-\langle\mathbf{b},\mathbf{a}'\rangle_E,$$ 
 where $\langle\cdot,\cdot\rangle_E$ denotes the usual Euclidean inner product over $\F_q$. Since the symplectic form is alternating, every vector is orthogonal to itself.

For an $\F_q$-linear subspace $\cC\subseteq\F_q^{2n}$, its symplectic dual is $\cC^{\perp_S}:=\{\mathbf{x}\in\F_q^{2n}:\langle\mathbf{x},\mathbf{c}\rangle_S=0 \text{ for every }\mathbf{c}\in\cC.\}$ The symplectic form is nondegenerate, and hence $\dim_{\F_q}\cC+\dim_{\F_q}\cC^{\perp_S}=2n.$ We say that $\cC$ is \emph{symplectic self-orthogonal} if $\cC\subseteq\cC^{\perp_S}$ and \emph{symplectic self-dual} if $\cC=\cC^{\perp_S}.$ For a vector $\mathbf{v}=(a_1,\ldots,a_n\mid b_1,\ldots,b_n)\in\F_q^{2n}$, define $\mathrm{wt}_S(\mathbf{v})=\bigl|{i\in[n]:(a_i,b_i)\neq(0,0)}\bigr|.$ For $\mathbf{u},\mathbf{v}\in\F_q^{2n}$, their symplectic distance is $d_S(\mathbf{u},\mathbf{v}):= \mathrm{wt}_S(\mathbf{u}-\mathbf{v}),$ and the symplectic minimum distance of a linear code $\cC\subseteq\F_q^{2n}$ is $d_S(\cC):=\min_{\mathbf{0}\neq\mathbf{v}\in\cC}\mathrm{wt}_S(\mathbf{v}).$

\begin{defn}[Symplectic Ball]
For an integer $d\geq 0$, define the $q$-ary symplectic ball 
$$B_q^S(2n,d):=\{\mathbf{x}\in\F_q^{2n}:\mathrm{wt}_S(\mathbf{x})\leq d\}.$$ Let $\Vol_q^S(2n,d):=|B_q^S(2n,d)|=\sum_{i=0}^{d}\binom{n}{i}(q^2-1)^i$ be its volume and for every fixed $0<\delta<1-1/q^2$, it holds 
\begin{equation}\label{eq:ball_volumn}
q^{2H_{q^2}(\delta)n-o(n)}\leq \Vol_q^S(2n,\lfloor\delta n\rfloor)\leq q^{2H_{q^2}(\delta)n}.\end{equation}
\end{defn}

The following standard connection between symplectic self-orthogonal codes and quantum stabilizer codes will be used throughout the paper.

\begin{lemma}[{\cite[Theorem~13]{ketkar2006nonbinary}}]
\label{lem:construct-quantum-codes}
Let $\cC\subseteq\F_q^{2n}$ be a $k$-dimensional symplectic self-orthogonal code. Then there exists a $q$-ary quantum code $[[n,n-k,d]]_q$ with $d\geq d_S(\cC^{\perp_S}).$
\end{lemma}

For completeness, we recall two standard upper bounds for quantum codes.

\begin{lemma}[{\cite[Lemma~1.2]{feng2004finite}}]
\label{lem:quantum-hamming}
For any pure stabilizer quantum code $[[n,k,d]]_q$, we have
\[
q^{n-k}
\geq
\sum_{i=0}^{\lfloor(d-1)/2\rfloor}
\binom{n}{i}(q^2-1)^i.
\]
\end{lemma}

\begin{lemma}[{\cite[Quantum Singleton Bound]{rains1999nonbinary}}]
\label{lem:quantum-singleton}
For every quantum code $[[n,k,d]]_q$, we have $n\geq k+2d-2.$
\end{lemma}

We next recall two finite length existence results for quantum codes that are relevant for comparison with our main theorem. Feng and Ma~\cite{feng2004finite} obtained a finite Gilbert–Varshamov-type bound for pure stabilizer quantum codes using $\F_{q^2}$-linear Hermitian self-orthogonal codes.

\begin{lemma}[Theorem~1.4,~\cite{feng2004finite}]\label{lem:feng-GV-bound}
Suppose that $n-2>k\ge 2$, $d\ge 2$, and $k$ is even. Then there exists a pure stabilizer quantum code $[[n,\,n-k,\,d]]_q$ provided that
\[
\sum_{i=0}^{d-1}\binom{n}{i}(q^2-1)^i<q^{k+2}.
\]
\end{lemma}

We finally recall the best finite length multiplicative improvement of the quantum GV bound that will serve as a comparison with our main result. A recent result gives an $\Theta(\sqrt n)$-multiplicative improvement over the standard quantum GV bound.

\begin{lemma}[{Corollary~5.8~\cite{yuan2026improvement}}]
\label{lem:previous-quantum-GV}
Fix a prime power $q$, a constant $0<\delta<1-\frac{1}{q^2}$ and let $d-1=\lfloor\delta n\rfloor$. There exists a constant $c_{q,\delta}>0$ such that, for all sufficiently large $n$, a $q$-ary quantum code $[[n,n-k,d]]_q$ exists whenever
\[
\frac{q^{2n-k}-1}{q-1}
<
c_{q,\delta}\sqrt n\,
\frac{q^{2n}}
{\sum_{i=0}^{d-1}\binom{n}{i}(q^2-1)^i}.
\]
\end{lemma}

\begin{rmk}
\label{rmk:previous-quantum-GV}
Lemma~\ref{lem:previous-quantum-GV} gives a multiplicative improvement of $\Theta(\sqrt{n})$ over the standard quantum GV bound. The main quantum result of this paper replaces the multiplicative $\sqrt n$ by $n$.
\end{rmk}

\section{Improvement of the Gilbert–Varshamov Bound for Linear codes}\label{sec:3}
 Throughout this section we assume $d-1=\lfloor \delta n\rfloor$ where $0<\delta<1-\frac{1}{q}.$ Let $M=|\mathbb{P}^{k}(\F_q)|=\frac{q^{k+1}-1}{q-1}$ and $a=\frac{\Vol_q(n,d-1)}{q^n}$. Let $G$ be a generator matrix for a random linear code of dimension $k+1$. It is clear that $G$ is a $(k+1)\times n$ random matrix.

In this section, we will prove the following result.
\begin{thm}\label{thm:main}
Let $q$ be any prime power and $0<\delta<1-1/q$ be any constant and $d-1=\lfloor \delta n\rfloor$. Then, there exists a $q$-ary linear $[n,k,d]$ code for all sufficiently large $n$ if
\begin{equation}
\label{eq:main-condition}
    \frac{q^k-1}{q-1}<c_{q,\delta} n\,\frac{q^n}{\Vol_q(n,d-1)},
\end{equation}
for some constant $c_{q,\delta}$.
\end{thm}
To prove this theorem, we need a Lemma from \cite{yuan2026improvement} which says that if we draw $r$ uniformly random vectors $\mathbf{u}_1,\ldots,\mathbf{u}_r$ from a Hamming ball of small size, it is exponentially unlikely that the sum of these $r$ vectors also lies in this Hamming ball. 
\begin{lemma}[{\cite[Lemma~3.3]{yuan2026improvement}}]\label{lem:escape}
Fix the prime power $q$ and $d-1=\lfloor\delta n\rfloor$ where $0<\delta<1-1/q$. There exists a constant $\gamma=O_{q,\delta}(1)>0$ such that, for all sufficiently large $n$ and every integer $2\le r\le n$, if the vectors
\[
    \mathbf{u}_1,\ldots,\mathbf{u}_r \overset{\mathrm{i.i.d.}}{\sim} \mathrm{Unif}(B_q(n,d-1)),
\]
then $\Pr[\mathbf{u}_1+\cdots+\mathbf{u}_r\in B_q(n,d-1)]\le e^{-\gamma n}.$
\end{lemma}

We can generalize the sum to any linear combination by a simple observation that $\mathbf{u}\in B_q(n,d-1)$ implies $\lambda \mathbf{u}\in B_q(n,d-1)$ for any scaler $\lambda\in \F_q^*$.
\begin{cor}\label{cor:coefficients}
Under the assumptions of Lemma~\ref{lem:escape}, for every $\lambda_1,\ldots,\lambda_r\in \F_q^*$,
\[
    \Pr[\lambda_1\mathbf{u}_1+\cdots+\lambda_r\mathbf{u}_r\in B_q(n,d-1)] \le e^{-\gamma n}.
\]
\end{cor}

\begin{defn}[Projective independence]
Let $T=\{[\mathbf{x}_1],\ldots,[\mathbf{x}_t]\}\subseteq \mathbb P^k(\mathbb F_q).$ $T$ is projectively independent if $\mathbf{x}_1,\ldots,\mathbf{x}_t$ are linearly independent in $\mathbb F_q^{k+1}$ and otherwise, $T$ is projectively dependent. A projectively dependent set is called minimally projectively dependent if every proper subset is projectively independent. 
\end{defn}

\begin{rmk}
It is clear that the projective independence does not depend on the representive element of $[\mathbf{x}_1],\ldots,[\mathbf{x}_t]$. Moreover, if $[\mathbf{x}_1],\ldots,[\mathbf{x}_t]$ is projectively dependent, then it always contains a minimally projectively dependent subset $\{[\mathbf{x}_{i_1}],\ldots,[\mathbf{x}_{i_r}],[\mathbf{y}]\}$ for some $2\leq r\leq k+1$ such that  
 $\mathbf{x}_{i_1},\ldots,\mathbf{x}_{i_r}$ are linearly independent and 
\begin{equation}\label{eq:minimal-relation}  \mathbf{y}=\lambda_1\mathbf{x}_{i_1}+\cdots+\lambda_r\mathbf{x}_{i_r}. 
\end{equation}
for some $\lambda_1,\ldots,\lambda_r\in\mathbb F_q^*$. 
\end{rmk}

\begin{lemma}\label{lem:fixed-circuit}
Let $\{[\mathbf{x}_1],\ldots,[\mathbf{x}_r],[\mathbf{y}]\}\subseteq\mathbb P^k(\mathbb F_q)$ be a mimimally projectively dependent set satisfying~\eqref{eq:minimal-relation}. Then, the probability that random vectors $\mathbf{x}_1G,\ldots,\mathbf{x}_rG, \mathbf{y}G$ all lie in $B_q(n,d-1)$ is at most $a^r\cdot e^{-\gamma n}.$
\end{lemma}

\begin{proof}
Since $\mathbf{x}_1,\ldots,\mathbf{x}_r$ are linearly independent, the random vectors $\mathbf{x}_1G,\ldots,\mathbf{x}_rG$ are independent and uniform in $\F_q^n$. The probability that they all lie in $B_q(n,d-1)$ is $a^r$. Conditioning on this event, we proceed to bound the probability that $\mathbf{y}G\in B_q(n,d-1)$. Since $\mathbf{y}G=\lambda_1\mathbf{x}_1G+\cdots+\lambda_r\mathbf{x}_rG$ and Corollary~\ref{cor:coefficients} shows that $\mathbf{y}G\in B_q(n,d-1)$ holds with probability at most $e^{-\gamma n}$.
\end{proof}

Given a projectively independent $r$-set $\{[\mathbf{x}_1],\ldots,[\mathbf{x}_r]\},$ \eqref{eq:minimal-relation} shows that there are at most $(q-1)^r$'s  $\mathbf{y}$ makes $\{[\mathbf{x}_1],\ldots,[\mathbf{x}_r],[\mathbf{y}]\}$ a minimally projectively dependent set $\mathbf{y}=\lambda_1\mathbf{x}_1+\cdots+\lambda_r\mathbf{x}_r$ with $\lambda_i\in \F_q^*$. This can be further reduced to $(q-1)^{r-1}$ when we only count the number of distinct $[\mathbf{y}]$.

The number of projectively independent $r$-sets is at most $\binom{M}{r}$. Every minimally projectively dependent set of size $r+1$ consists of a projectively indpendent $r$-sets $[\mathbf{x}_1],\ldots,[\mathbf{x}_r]$ and $[\mathbf{y}]$ meeting \eqref{eq:minimal-relation}.
Therefore, the total number of such sets is at most $\binom Mr (q-1)^{r-1}$. Recall the definition of $S(G)$ in \eqref{eqn:S_G}.

\begin{lemma}\label{lem:independence}
Using the above notations, if $Ma\le \frac{\gamma}{2(q-1)}n,$ then
\[
    \Pr[S(G)\text{ is projectively dependent}]\le \frac{1}{q-1}e^{-\gamma n/2}.
\]
\end{lemma}

\begin{proof}
If $S(G)$ is projectively dependent, then it must contain a minimally projectively dependent subset. Thus, it suffice to consider the minimally projectively dependent set. Taking a union bound over all minimally projectively dependent set of size $r+1$ for $r=2,\ldots,k+1$, by Lemma~\ref{lem:fixed-circuit} we have
\begin{align*}
    \Pr[S(G)\text{ is projectively dependent}]&\le\sum_{r=2}^{k+1} \binom Mr (q-1)^{r-1}a^r e^{-\gamma n}
    =\frac{e^{-\gamma n}}{q-1} \sum_{r=2}^{k+1} \binom Mr\bigl((q-1)a\bigr)^r\\
    &\le\frac{e^{-\gamma n}}{q-1} \bigl(1+(q-1)a\bigr)^M \le \frac{1}{q-1}e^{\bigl(-\gamma n+(q-1)Ma\bigr)} \\
    &\le \frac{1}{q-1}e^{-\gamma n/2}.
    \end{align*}
\end{proof}

\begin{lemma}\label{lem:hyperplane}
Let $G\in \F_q^{(k+1)\times n}$ be a matrix and suppose that $S(G)$ is projectively independent. Then, there exists a full-rank matrix $U\in \F_q^{k\times (k+1)}$ such that $S(UG)$ is empty.
\end{lemma}

\begin{proof}
Let $S(G)=\{[\mathbf{s}_1],\ldots,[\mathbf{s}_t]\}$. 
Since $S(G)$ are projectively independent, $\mathbf{s}_1,\ldots,\mathbf{s}_t$ are linearly independent with $t\leq k+1$ and we can extend them to a basis of $\F_q^{k+1}$. Let $\mathbf{s}_1,\ldots,\mathbf{s}_{k+1}$ be a basis of $\F_q^{k+1}$. Define a $\F_q$-linear map $\varphi:\F_q^{k+1}\to\F_q$ such that $\varphi(\mathbf{s}_i)=1$ for $i\in [t]$ and $\varphi(\mathbf{s}_i)=0$ for $i\in [t+1, k+1]$. 
Note that $\varphi$ is a non-zero map and thus $\ker(\varphi)$ is a subspace of dimension $k$.

Let $\mathbf{u}_1,\ldots,\mathbf{u}_k\in \F_q^{k+1}$ be a basis of $\ker(\varphi)$ and let $U\in \F_q^{k\times (k+1)}$ be a matrix whose rows are $\mathbf{u}_1,\ldots,\mathbf{u}_k$. Then, we have 
$$
S(UG)=\{[\mathbf{x}]\in\mathbb{P}^{k-1}(\F_q) : \mathbf{x}UG\in B_q(n,d-1)\}.
$$
Assume that $S(UG)$ is nonempty and then we can find a projectively independent set $\{[\mathbf{x}_1],\ldots,[\mathbf{x}_r]\}\subseteq S(UG)$. It follows that 
$\{[\mathbf{x}_1U],\ldots,[\mathbf{x}_rU]\}\subseteq S(G)$ are projectively independent set. However, by the construction of linear map $\varphi$ and $U$, we have $\varphi(\mathbf{x})\neq 0$ for $[\mathbf{x}]\in S(G)$ and $\varphi(\mathbf{x}U)=0$ for any $\mathbf{x}\in \F_q^k$. A contradiction happens and we complete the proof.

\end{proof}

\begin{proof}[Proof of Theorem~\ref{thm:main}]
Let $\gamma=O_{q,\delta}(1)$ be the constant from Lemma~\ref{lem:escape}, and choose $c_{q,\delta}:=\frac{\gamma}{4q(q-1)}.$ By condition~\eqref{eq:main-condition}, it holds 
$$Ma=\frac{q^{k+1}-1}{q-1}a< \frac{\gamma}{4(q-1)}n+a\leq \frac{\gamma}{2(q-1)}n$$
as $a\leq 1$ and $n$ is large enough. Lemma~\ref{lem:independence} shows 
$$\Pr[S(G)\text{ is projectively dependent}]\le \frac{1}{q-1}e^{-\gamma n/2}<1.$$ Hence, there exists a matrix $G\in\mathbb F_q^{(k+1)\times n}$ such that $S(G)$ is projectively independent. Lemma~\ref{lem:hyperplane} further shows that there exists $UG\in \mathbb F_q^{(k)\times n}$ with $S(UG)=\emptyset$. Let $\cC$ be a linear code with generator matrix $UG$. Then, $\cC$ is a q-ary $[n,k,d]$-linear code.
\end{proof}

\section{Improvement of the Gilbert-Varshamov Bound for Quantum Codes}\label{sec:4}
In this section, we assume throughout that $d-1=\lfloor \delta n\rfloor$ and $H_{q^2}(\delta)<\frac{1}{2}.$ Our main theorem is the following.

\begin{thm}\label{thm:quantum-gv-linear-factor}
Let $q$ be a prime power and $\delta\in (0,1-1/q^2)$ with $H_{q^2}(\delta)<\frac{1}{2}.$ Let $d-1=\lfloor\delta n\rfloor$ and  integer $1\leq k\leq n$. There exists a $k$-dimensional symplectic self-orthogonal code $\cC\subseteq \F_q^{2n}$ with $d_S(\cC^{\perp_S})\geq d$ for large enough $n$ if 
\begin{equation}\label{eq:quantum-gv-linear-factor}
\frac{q^{2n-k}-1}{q-1}<c_{q,\delta}\, n\,\frac{q^{2n}}{\mathrm{Vol}_q^S(2n,d-1)},
\end{equation}
for some constant $c_{q,\delta}$.
\end{thm}

\begin{rmk}\label{rem:quantum-gv-range}
The condition $H_{q^2}(\delta)<\frac{1}{2}$ is necessary for our $\Theta(n)$ improvement to be applied. This is because in our symplectic construction, we require $k\leq n$.
\end{rmk}

Choose uniformly at random a $k-1$-dimensional symplectic self-orthogonal subspace $\cC\subseteq \mathbb F_q^{2n}$ satisfying $\cC\subseteq \cC^{\perp_S}$. Since $\dim \cC=k-1$, we have $\dim \cC^{\perp_S}=2n-k+1.$ 


\begin{defn}[$\cC$-equivalence]\label{def:bad-direction}
Let $V=\cC^{\perp_S}/\cC$ be a quotient space. We denote by $[\mathbf{x}]_{\cC}$ a $\cC$-equivalence class in the equivalence class $V\setminus \{\mathbf{0}\}\slash \F_q^*$. $[\mathbf{x}]_{\cC}$ is said to be bad if there exists a $\mathbf{v}\in [\mathbf{x}]_{\cC}$ such that $\mathbf{v}\in \bigl(\cC^{\perp_S}\cap B_q^S(2n,d-1)\bigr)\slash \cC$. Let $\mathcal{B}(\cC)$ be the collection of all bad $\cC$-equivalence classes.
\end{defn}

\begin{defn}[$\cC$-independence]\label{def:C-independence}
Let $[\mathbf{v}_1]_{\cC},\ldots,[\mathbf{v}_t]_{\cC}$ be distinct $\cC$-equivalence classes. We say that $[\mathbf{v}_1]_{\cC},\ldots,[\mathbf{v}_t]_{\cC}$ are \emph{$\cC$-independent} if $\lambda_1\mathbf{v}_1+\cdots+\lambda_t\mathbf{v}_t\in \cC$ implies that $\lambda_1=\cdots=\lambda_t=0$ for all $\lambda_1,\ldots,\lambda_t\in\F_q$. Otherwise, they are called \emph{$\cC$-dependent}. A $\cC$-dependent set is called \emph{minimally $\cC$-dependent} set if every proper subset of it is $\cC$-independent. 
\end{defn}
\begin{rmk}
It can be shown that the $\cC$-independence of $[\mathbf{v}_1]_{\cC},\ldots,[\mathbf{v}_t]_{\cC}$ does not depend on the choice of the representative element $\mathbf{v}_1,\ldots,\mathbf{v}_t$.
\end{rmk}


Instead of requiring $\cC^{\perp_S}\cap B_q^S(2n,d-1)=\{\mathbf{0}\},$ we shall prove that there exists a symplectic self-orthogonal code $\cC$ satisfying 
\[
\cC\cap B_q^S(2n,d-1)=\{\mathbf{0}\},
\qquad
\mathcal{B}(\cC)\text{ is $\cC$-independent}.
\]

\begin{lemma}\label{lem:one-step-elimination}
Suppose that $\cC\subseteq\mathbb F_q^{2n}$ is a $(k-1)$-dimensional symplectic self-orthogonal code satisfying
\[
\cC\cap B_q^S(2n,d-1)=\{\mathbf{0}\},
\qquad
\mathcal B(\cC)\text{ is $\cC$-independent}.
\]
Then $\cC$ can be extended to a $k$-dimensional symplectic self-orthogonal code $\cC'$ such that $\cC'^{\perp_S}\cap B_q^S(2n,d-1)=\{\mathbf{0}\}.$
\end{lemma}

\begin{proof}
Assume $\mathcal{B}(\cC)=\{[\mathbf{v}_1]_{\cC},\ldots,[\mathbf{v}_t]_{\cC}\}$, where $\mathbf{v}_i\in B_q^S(2n,d-1)$ is a representative of $[\mathbf{v}_i]_{\cC}$ for $1\le i\le t$. If $t=0$, then $\cC^{\perp_S}\cap B_q^S(2n,d-1)=\{\mathbf{0}\}$ and there is nothing to prove. Otherwise,
define the linear map
\[
\varphi:\cC^{\perp_S}\longrightarrow\mathbb F_q^t,
\qquad
\varphi(\mathbf{z})
=
\bigl(
\langle\mathbf{v}_1,\mathbf{z}\rangle_S,\ldots,
\langle\mathbf{v}_t,\mathbf{z}\rangle_S
\bigr).
\]
We prove that $\varphi$ is surjective. Otherwise, there exists not all zeros $\lambda_1,\ldots,\lambda_t\in\mathbb F_q$ such that $$0=\sum_{i=1}^t\lambda_i\langle\mathbf{v}_i,\mathbf{z}\rangle_S=\langle\sum_{i=1}^t\lambda_i\mathbf{v}_i,\mathbf{z}\rangle_S$$ 
for any $\mathbf{z}\in\cC^{\perp_S}$. This implies that $\sum_{i=1}^t \lambda_i \mathbf{v}_i\in \cC$. This is impossible as $[\mathbf{v}_1]_{\cC},\ldots,[\mathbf{v}_t]_{\cC}$ are $\cC$-independent. Thus,  $\mathrm{rank}(\varphi)=t=\dim\mathbb F_q^t$ and  $\varphi$ is surjective.

Now we can find $\mathbf{z}\in\cC^{\perp_S}$ such that $\varphi(\mathbf{z})=(1,\ldots,1).$ Note that $\mathbf{z}\notin\cC$ since every vector in $\cC$ is symplectically orthogonal to  $\mathbf{v}_i\in\cC^{\perp_S}$, for $i\in [t]$. Set $\cC':=\cC+\langle\mathbf{z}\rangle.$ Then $\dim\cC'=k$. Moreover, since $\mathbf{z}\in\cC^{\perp_S}$ and $\langle\mathbf{z},\mathbf{z}\rangle_S=0$, the subspace $\cC'$ is symplectic self-orthogonal, and $\cC'^{\perp_S}\subseteq \cC^{\perp_S}.$

It remains to show $\cC'^{\perp_S}\cap B_q^S(2n,d-1)=\{\mathbf{0}\}$. Suppose for contradiction that there exists $\mathbf{0}\ne\mathbf{x}\in\cC'^{\perp_S}\cap B_q^S(2n,d-1).$ Since $\cC\cap B_q^S(2n,d-1)=\{\mathbf{0}\}$, we have $\mathbf{x}\notin\cC$. Therefore, $[\mathbf{x}]_{\cC}$ is a bad $\cC$-equivalence class, and hence $[\mathbf{x}]_{\cC}=[\mathbf{v}_i]_{\cC}$ for some $i\in\{1,\ldots,t\}$. It follows that there exist $\mu\in\mathbb F_q^*$ and $\mathbf{c}\in\cC$ such that $\mathbf{x}=\mu\mathbf{v}_i+\mathbf{c}.$ On the other hand,  $\mathbf{x}\in\cC'^{\perp_S}$ implies $\langle\mathbf{x},\mathbf{z}\rangle_S=0$.  However, $\langle\mathbf{x},\mathbf{z}\rangle_S=\mu\langle\mathbf{v}_i,\mathbf{z}\rangle_S+\langle\mathbf{c},\mathbf{z}\rangle_S=\mu,$ and a contradiction happens. The proof is completed.
\end{proof}

\begin{lemma}\label{lem:symplectic-incidence}
Let $\cC$ be chosen uniformly at random from all $(k-1)$-dimensional symplectic self orthogonal subspaces of $\mathbb F_q^{2n}$. Given any $r$-dimensional linear subspace $U\subseteq\mathbb F_q^{2n}$, we have $\Pr\!\left[U\subseteq \cC^{\perp_S},\;U\cap\cC=\{\mathbf{0}\}\right]\le q^{-(k-1)r}.$
\end{lemma}

\begin{proof}
Let $\cC_0=\{\mathbf{0}\}$. For $i=0,\ldots,k-2$, choose $\mathbf{z}_{i+1}$ uniformly at random from $\cC_i^{\perp_S}\setminus\cC_i$ and set $\cC_{i+1}:=\cC_i+\langle\mathbf{z}_{i+1}\rangle$. Every $\cC_i$ is an $i$-dimensional symplectic self orthogonal subspace. Moreover, every ordered basis is chosen with the same probability, and every $(k-1)$-dimensional subspace has the same number of ordered bases. Thus, $\cC_{k-1}$ is uniformly distributed as $\cC$.

Suppose that $\cC_i$ satisfies $U\subseteq\cC_i^{\perp_S}$ and $U\cap\cC_i=\{\mathbf{0}\}$. Then $\dim(\cC_i+U)=i+r$. For $\cC_{i+1}$ to satisfy the same two conditions, we must have $\mathbf{z}_{i+1}\in (\cC_i+U)^{\perp_S}\setminus\cC_i$. Since $\cC_i\subseteq(\cC_i+U)^{\perp_S}$, there are $q^{2n-i-r}-q^i$ such choices, while $\cC_i^{\perp_S}\setminus\cC_i$ contains $q^{2n-i}-q^i$ vectors. Therefore, the probability that the next subspace still satisfies both conditions is at most $\frac{q^{2n-i-r}-q^i}{q^{2n-i}-q^i}\le q^{-r}.$

If $\cC_{k-1}$ satisfies both conditions, then so
does every $\cC_i$. Multiplying these bounds for
$i=0,\ldots,k-2$ gives
\[
\Pr\!\left[
U\subseteq\cC^{\perp_S},\;
U\cap\cC=\{\mathbf{0}\}
\right]
\le(q^{-r})^{k-1}=q^{-(k-1)r}.
\]
The proof is completed.
\end{proof}

The following lemma will be used to bound the probability that a collection of low-weight vectors is linearly independent, but its span has a nontrivial intersection with the random subspace $\cC$. This situation is the major challenge to be overcome compared to the Hamming case.

\begin{lemma}\label{lem:quotient-dependence}
Let $\cC$ be chosen uniformly at random from all $(k-1)$-dimensional symplectic self-orthogonal subspaces of $\F_q^{2n}$. Let $\mathbf{v}_1,\ldots,\mathbf{v}_t\in\F_q^{2n}$ be $\F_q$-linearly independent vectors. Set $U:=\mathrm{span}\{\mathbf{v}_1,\ldots,\mathbf{v}_t\}$ and $\mathbf{c}:=\sum_{i=1}^t\lambda_i\mathbf{v}_i,$ where $\lambda_1,\ldots,\lambda_t\in\F_q^\ast$. Then
\begin{equation}\label{eq:quotient-dependence}
\Pr_{\cC}\left[
U\subseteq\cC^{\perp_S},\;
U\cap\cC=\langle\mathbf{c}\rangle
\right]
\le q^{-(k-1)t}\cdot q^{-2(n-k+1)+t}.
\end{equation}
\end{lemma}

\begin{proof}
If $\langle \mathbf{c},\mathbf{v}_i\rangle_S\neq 0$ for some $i$, then $\mathbf{c}\notin \cC$ as $\mathbf{v}_i\in \cC^{\perp_S}$. In this case, the probability is zero and \eqref{eq:quotient-dependence} holds trivially. Then, we assume that $\langle \mathbf{c},\mathbf{v}_i\rangle_S=0$ for $1\leq i\leq t$.

We first bound the probability that $\mathbf{c}\in \cC$. Due to the symplectic structure, we have
\begin{equation}\label{eq:c-in-C0}
\Pr[\mathbf{c}\in \cC]=\frac{q^{k-1}-1}{q^{2n}-1}.
\end{equation}

Conditioning on $\mathbf{c}\in \cC$, $\cC$ is uniformly distributed among the $(k-1)$-dimensional symplectic self-orthogonal subspaces containing $\mathbf{c}$. Let $V_{\mathbf{c}} := \langle \mathbf{c} \rangle$. Then $\cC' := \cC/V_\mathbf{c}$ is a uniformly random $(k-2)$-dimensional symplectic self-orthogonal subspace of the symplectic quotient $V_{\mathbf{c}}^{\perp_S}/V_{\mathbf{c}}$.

Since $\mathbf{c}\in U$ and $\langle\mathbf{c},\mathbf{v}_i\rangle_S=0$ for all $i\in[t]$, we have $U\subseteq V_{\mathbf{c}}^{\perp_S}$. Let $U'$ denote the image of $U$ in $V_{\mathbf{c}}^{\perp_S}/V_{\mathbf{c}}$. Then $\dim U'=t-1$. It is clear that conditioning on $\mathbf{c}\in\cC$,
\[
\begin{aligned}
U\subseteq\cC^{\perp_S},\quad
U\cap\cC=V_{\mathbf{c}}\quad\Longleftrightarrow\quad
U'\subseteq\cC'^{\perp_S},\quad
U'\cap\cC'=\{\mathbf{0}\}.
\end{aligned}
\]
Therefore,
\begin{equation}\label{eq:conditional-quotient-dependence}
\Pr_{\cC}\left[
U\subseteq\cC^{\perp_S},
U\cap\cC=V_{\mathbf{c}}
\mid\mathbf{c}\in\cC
\right]=
\Pr_{\cC'}\left[
U'\subseteq\cC'^{\perp_S},
U'\cap\cC'=\{\mathbf{0}\}
\right].
\end{equation}
Lemma~\ref{lem:symplectic-incidence} shows that this probability is at most $q^{-(k-2)(t-1)}$. Combining~\eqref{eq:c-in-C0} and~\eqref{eq:conditional-quotient-dependence}, we obtain
\[
\begin{aligned}
\Pr_{\cC}\left[
U\subseteq\cC^{\perp_S},
\ U\cap\cC=V_{\mathbf{c}}
\right]&=
\Pr[\mathbf{c}\in\cC]\,
\Pr_{\cC'}\left[
U'\subseteq\cC'^{\perp_S},
\ U'\cap\cC'=\{\mathbf{0}\}
\right]\\
&\le
\frac{q^{k-1}-1}{q^{2n}-1}
\cdot q^{-(k-2)(t-1)}\le
q^{-(k-1)t}\cdot q^{-2(n-k+1)+t}.
\end{aligned}
\]
The proof is completed.
\end{proof}

Similar to the linear codes, we can bound the probability that the sum of random low weight vectors is still a low weight vectors. 
\begin{lemma}[{\cite[Lemma~5.5]{yuan2026improvement}}]\label{lem:symplectic-ball-escape}
Let $q$ be a prime power and $d-1=\lfloor\delta n\rfloor$ with $0<\delta<1-1/q^2$. Then, there exists a constant $h=O_{q,\delta}(1)$ for all sufficiently large $n$ and every integer $2\le r\le n$, if
\[
\mathbf{u}_1,\ldots,\mathbf{u}_r\stackrel{\mathrm{i.i.d.}}{\sim}\mathrm{Unif}\!\left(B_q^S(2n,d-1)\right),
\]
then
\[
\Pr\!\left[\mathbf{u}_1+\cdots+\mathbf{u}_r\in B_q^S(2n,d-1)\right]\le e^{-hn}.
\] 
\end{lemma}

\begin{cor}\label{cor:symplectic-ball-linear-combination}
Under the assumptions in Lemma~\ref{lem:symplectic-ball-escape}, for every $\lambda_1,\ldots,\lambda_r\in\mathbb F_q^\ast$, we have
\[
\Pr\!\left[
\lambda_1\mathbf{u}_1+\cdots+\lambda_r\mathbf{u}_r
\in B_q^S(2n,d-1)
\right]
\le e^{-hn}.
\]
\end{cor}

Suppose that $\mathcal B(\cC)$ is $\cC$-dependent. Then it must contain a minimally $\cC$-dependent subset $[\mathbf{v}_1]_{\cC},\ldots,[\mathbf{v}_t]_{\cC}$ with $t\geq 3$. We next bound the probability that such a minimally $\cC$-dependent subset exists. 
The minimal $\cC$-dependence set yields nonzero coefficients $\lambda_1,\ldots,\lambda_t$ such that $\mathbf{c}:=\sum_{i=1}^t\lambda_i\mathbf{v}_i\in\cC.$

We divide our discussion into two cases. If $\mathbf{c}=\mathbf{0}$, then the representatives themselves are linearly dependent. This is an analogue case can be found in linear code argument. If $\mathbf{c}\neq \mathbf{0}$, the subspace spanned by $\mathbf{v}_1,\ldots,\mathbf{v}_t$ has a nontrivial intersection with $\cC$. This is unlikely to happen by Lemma~\ref{lem:quotient-dependence}. The following lemma bound the probability that $\mathcal B(\cC)$ is $\cC$-dependent.

\begin{lemma}\label{lem:minimal-bad-directions}
Let $E_t$ be the event that $\mathcal B(\cC)$ contains a minimally $\cC$-dependent subset of size $t\geq 3$. Then, 
\begin{equation}\label{eq:minimal-bad-bound}
\Pr[E_t]\le \frac{(q-1)^{t-1}}{t!} [e^{-hn}( \frac{\Vol_q^S(2n,d-1)}{q^{k-1}} )^{t-1}+ q^{-2(n-k+1)+t}( \frac{\Vol_q^S(2n,d-1)}{q^{k-1}})^t ].
\end{equation}
\end{lemma}

\begin{proof}
    Let $\{[\mathbf v_1]_{\cC},\ldots,[\mathbf v_t]_{\cC}\}\subseteq \mathcal{B}(\cC)$ be the minimally $\cC$-dependent set. Due to the definition of $\mathcal{B}(\cC)$, without loss of generality, we assume that $\mathbf v_i\in B_q^S(2n,d-1)$ for $i\in [t]$. Moreover, the minimal $\cC$-dependence implies that there exist $\lambda_1,\ldots,\lambda_t\in\mathbb F_q^*$ such that $\sum_{i=1}^t\lambda_i\mathbf v_i\in \cC$. 
    In other words, we can write $\mathbf v_t = \sum_{i=1}^{t-1}\alpha_i\mathbf v_i+\mathbf c$ for some $\mathbf c\in \cC$ and $\alpha_1,\ldots,\alpha_{t-1}\in \F_q^*$. 

    Let $R_t(\cC)$ be the number of ordered tuples $(\mathbf{v}_1,\ldots,\mathbf{v}_t)$ with $\mathbf{v}_i\in\cC^{\perp_S}\cap B_q^S(2n,d-1)$ for $i\in[t]$, such that $[\mathbf{v}_1]_{\cC},\ldots,[\mathbf{v}_{t-1}]_{\cC}$ are $\cC$-independent and $\mathbf{v}_t-\sum_{i=1}^{t-1}\alpha_i\mathbf{v}_i\in\cC$ for some $\alpha_1,\ldots,\alpha_{t-1}\in\mathbb{F}_q^*$. For each such tuple, write $\mathbf{c}:=\mathbf{v}_t-\sum_{i=1}^{t-1}\alpha_i\mathbf{v}_i\in\cC$. Since $[\mathbf{v}_1]_{\cC},\ldots, [\mathbf{v}_{t-1}]_{\cC}$ are $\cC$-independent, we have
    \[
    U\cap\cC=\{\mathbf{0}\},
    \quad U:=\mathrm{span}\{\mathbf{v}_1,\ldots,\mathbf{v}_{t-1}\},\quad \dim U=t-1.
    \]
    We divide our discussion into two cases $\mathbf{c}=\mathbf{0}$ and $\mathbf{c}\neq\mathbf{0}$.

    We first consider the case $\mathbf{c}=\mathbf{0}$.
    Fix $(\alpha_1,\ldots,\alpha_{t-1})\in(\mathbb F_q^*)^{t-1}$ and write $\mathbf x := \sum_{i=1}^{t-1}\alpha_i\mathbf v_i.$  By Corollary~\ref{cor:symplectic-ball-linear-combination}, the number of ordered $(t-1)$-tuples $(\mathbf v_1,\ldots,\mathbf v_{t-1}) \in B_q^S(2n,d-1)^{t-1}$ such that $\mathbf x\in B_q^S(2n,d-1)$ is at most $e^{-hn}\Vol_q^S(2n,d-1)^{t-1}.$  Fixing a linearly independent tuple $(\mathbf{v}_1,\ldots,\mathbf{v}_{t-1})$, its contribution to $R_t(\cC)$ requires $U\subseteq\cC^{\perp_S}$ and $U\cap\cC=\{\mathbf{0}\}$. By Lemma~\ref{lem:symplectic-incidence}, we have $\Pr_{\cC}\!\left[U\subseteq\cC^{\perp_S},\ U\cap\cC=\{\mathbf{0}\}\right]\le q^{-(k-1)(t-1)}.$ This implies that the expected number of tuples counted by $R_t(\cC)$ with $\mathbf{c}=\mathbf{0}$ and fixed coefficients is at most $e^{-hn}\left( \frac{\Vol_q^S(2n,d-1)}{q^{k-1}} \right)^{t-1}$.
    
    We now proceed to the case $\mathbf{c}\neq\mathbf{0}$. Since $\mathbf{c}\in\cC$ and $U\cap\cC=\{\mathbf{0}\}$, we have $\mathrm{span}\{\mathbf{v}_1,\ldots,\mathbf{v}_t\}=U\oplus\langle\mathbf{c}\rangle$ and $\mathrm{span}\{\mathbf{v}_1,\ldots,\mathbf{v}_t\}\cap\cC=\langle\mathbf{c}\rangle.$ In particular, $\mathbf{v}_1,\ldots,\mathbf{v}_t$ are linearly independent. For each fixed ordered $t$-tuple of low weight vectors $(\mathbf{v}_1,\ldots,\mathbf{v}_t)$, we apply Lemma~\ref{lem:quotient-dependence} to $\mathbf{c}=-\alpha_1\mathbf{v}_1-\cdots-\alpha_{t-1}\mathbf{v}_{t-1}+\mathbf{v}_t$ to obtain
    \[
    \Pr_{\cC}\!\left[\mathbf{v}_1,\ldots,\mathbf{v}_t\in\cC^{\perp_S},
\ \mathrm{span}\{\mathbf{v}_1,\ldots,
\mathbf{v}_t\}\cap\cC
=\langle\mathbf{c}\rangle
\right]
\le q^{-(k-1)t}q^{-2(n-k+1)+t}.
    \]
   
    There are at most $\Vol_q^S(2n,d-1)^t$ $t$-tuples $(\mathbf{v}_1,\ldots,\mathbf{v}_t)$. Since there are $(q-1)^{t-1}$ such $(\alpha_1,\ldots,\alpha_{t-1})$, the total expected number of tuples with $\mathbf{c}\neq\mathbf{0}$ is at most $(q-1)^{t-1}q^{-2(n-k+1)+t}\left(\frac{\Vol_q^S(2n,d-1)}{q^{k-1}}\right)^t.$

    Combining the contributions from $\mathbf{c}=\mathbf{0}$ and $\mathbf{c}\neq \mathbf{0}$, yields 
    \[ 
    \mathbb E_{\cC}[R_t(\cC)] \le (q-1)^{t-1}[e^{-hn}( \frac{\Vol_q^S(2n,d-1)}{q^{k-1}} )^{t-1}+ q^{-2(n-k+1)+t}( \frac{\Vol_q^S(2n,d-1)}{q^{k-1}})^t ]
    \]
    Note that we only need to count the number of minimal $\cC$-dependence sets instead of tuples, thus we have
     $\Pr[E_t]\leq \frac{\mathbb E_{\cC}[R_t(\cC)]}{t!}$ as each set will be counted at least $t!$ times in the tuples.
     The proof is completed.
\end{proof}

Let Lemma~\ref{lem:minimal-bad-directions} range from $t=3$ to $t=L$ gives the following result.

\begin{cor}\label{cor:minimal-dependence}
With the above notation, for every integer $3\leq L\leq n$, then
\[
\sum_{t=3}^L\Pr\bigl[ E_t\bigr] 
\le \left(e^{-hn}+q^{-2(n-k+1)}\right)e^{q(q-1)\Vol_q^S(2n,d-1)q^{-(k-1)}}.
\]
\end{cor}

\begin{proof}
   By Lemma~\ref{lem:minimal-bad-directions} and the union bound over $t=3,\ldots,L$, we have 
   \[
  \sum_{t=3}^L \Pr[E_t] \le e^{-hn} \sum_{t=3}^{L}\frac{1}{t!}(\frac{(q-1)\Vol_q^S(2n,d-1)}{q^{k-1}})^{t-1} + \frac{1}{q^{2(n-k+1)}(q-1)}\sum_{t=3}^{L}\frac{1}{t!} (\frac{ q(q-1)\Vol_q^S(2n,d-1)}{q^{k-1}})^t.
   \]
   For the first sum, 
   \[
   \sum_{t=3}^{L} \frac{ \bigl((q-1)\Vol_q^S(2n,d-1)q^{-(k-1)}\bigr)^{t-1} }{t!} \le e^{(q-1)\Vol_q^S(2n,d-1)q^{-(k-1)}} \le e^{q(q-1)\Vol_q^S(2n,d-1)q^{-(k-1)}}.
   \]
   For the second sum, we have 
   \[ 
  \frac{1}{q-1} \sum_{t=3}^{L} \frac{ \bigl(q(q-1)\Vol_q^S(2n,d-1)q^{-(k-1)}\bigr)^t }{t!} \le e^{q(q-1)\Vol_q^S(2n,d-1)q^{-(k-1)}}. 
   \]
   The proof is completed when we combine these two sums together.
\end{proof}

We are now ready to prove Theorem~\ref{thm:quantum-gv-linear-factor}.

\begin{proof}[Proof of Theorem~\ref{thm:quantum-gv-linear-factor}]
By the standard entropy upper bound for the symplectic Hamming ball, we have $\Vol_q^S(2n,d-1)\le q^{2H_{q^2}(\delta)n}$ for $d-1=\lfloor \delta n\rfloor$. Let $\cC$ be chosen uniformly at random from all $(k-1)$-dimensional symplectic self-orthogonal subspaces of $\mathbb F_q^{2n}$. Define $Z:=\left|\bigl(\cC^{\perp_S}\cap B_q^S(2n,d-1)\bigr)\setminus\{\mathbf{0}\}\right|,$ the number of nonzero low-weight vectors contained in $\cC^{\perp_S}$. For every fixed nonzero $\mathbf{v}\in\mathbb F_q^{2n}$, the symplectic structure implies that $\Pr_{\cC}[\mathbf{v}\in\cC^{\perp_S}]=\frac{q^{2n-k+1}-1}{q^{2n}-1}\le q^{-(k-1)}.$ By linearity of expectation,  
$$\mathbb E[Z]=\sum_{\mathbf{v}\in B_q^S(2n,d-1)\setminus \{\mathbf{0}\}} \Pr[\mathbf{v}\in \cC^{\perp_S}] \le \Vol_q^S(2n,d-1)\cdot q^{-(k-1)}$$

Under condition \eqref{eq:quantum-gv-linear-factor}, we have 
\[
\begin{aligned}
\frac{\Vol_q^S(2n,d-1)}{q^{k-1}}=q(q-1)(\frac{q^{2n-k}-1}{q-1}\cdot \frac{\Vol_q^S(2n,d-1)}{q^{2n}})+\frac{\Vol_q^S(2n,d-1)}{q^{2n-1}}<q(q-1)c_{q,\delta}n+1,
\end{aligned}
\]
where the last inequality follows from~\eqref{eq:quantum-gv-linear-factor}. Let $h=O_{q,\delta}(1)$ be the constant from Lemma~\ref{lem:symplectic-ball-escape}. Let $c_{q,\delta}>0$ such that $c_{q,\delta}<\min\{\frac{1-2H_{q^2}(\delta)}{64q(q-1)},\frac{\min\{h,(1-2H_{q^2}(\delta))\ln q\}}{2q^2(q-1)^2}\}.$

We divide our discussion into two cases.

\medskip
\noindent
\textbf{Case 1: $n-k+1<\frac{(1-2H_{q^2}(\delta)) n}{2}$.}

In this case, we have $2nH_{q^2}(\delta)\leq 2k-n-2$ and thus
$$\Vol_q^S(2n,d-1)\cdot q^{-(k-1)}\leq q^{2nH_{q^2}(\delta)-k+1}\leq q^{k-n-1}\leq \frac{1}{q^2}$$ 
as $k<n$.
Therefore, $\mathbb E[Z]\leq \frac{1}{q}.$ By Markov's inequality, we have $\Pr[Z>0]\le \mathbb E[Z]\leq \frac{1}{q}.$ Therefore, here exists $\cC$ satisfying $\cC^{\perp_S}\cap B_q^S(2n,d-1)=\{\mathbf{0}\}.$

Choose any $k$-dimensional symplectic self-orthogonal code $\cC'\supset \cC$. Since $\cC'^{\perp_S}\subseteq \cC^{\perp_S},$ we still have $\cC'^{\perp_S}\cap B_q^S(2n,d-1)=\{\mathbf{0}\}.$ Thus, $d_S(\cC'^{\perp_S})\ge d$, and the theorem holds in this case.

\medskip
\noindent
\textbf{Case 2: $n-k+1\ge \frac{(1-2H_{q^2}(\delta)) n}{2}$.}

Set $L:=\left\lfloor\frac{(1-2H_{q^2}(\delta))n}{8}\right\rfloor$.  
By Markov's inequality and the choice of $c_{q,\delta}$, we obtain 
$$\Pr[Z>L]\le\frac{\mathbb E[Z]}{L}<\frac{1}{4}$$ 
for all sufficiently large $n$.

It remains to consider the case that $Z\le L$. In this case, we have $|\mathcal B(\cC)|\le L$. Let $E$ be the event that $\mathcal B(\cC)$ contain a minimal $\cC$-dependent subset of size at most $L$.
By Corollary~\ref{cor:minimal-dependence}, we have 
\begin{equation}\label{eq:L-event}
\Pr[E]\leq (e^{-hn}+q^{-2(n-k+1)})e^{
q(q-1))\Vol_q^S(2n,d-1)q^{-(k-1)}}.
\end{equation}
Since $n-k+1\ge\frac{(1-2H_{q^2}(\delta))n}{2},$ we have $q^{-2(n-k+1)}\le e^{-(1-2H_{q^2}(\delta))\ln q\cdot n}$. The condition \eqref{eq:quantum-gv-linear-factor} implies 
$$q(q-1)\Vol_q^S(2n,d-1)\cdot q^{-(k-1)}<q^2(q-1)^2c_{q,\delta}n+O_q(1).$$ Thus,
\[
\Pr[E]\leq e^{-hn+q^2(q-1)^2c_{q,\delta}n+O(1)}+e^{-(1-2H_{q^2}(\delta))\ln q\cdot n+q^2(q-1)^2c_{q,\delta}n+O(1)}=e^{-\Omega(n)}.
\]

It remains to ensure that $\cC$ itself contains no nonzero low-weight vector. 
For every fixed nonzero $\mathbf{v}\in\mathbb F_q^{2n}$, the symplectic structure implies that  $\Pr[\mathbf{v}\in\cC]=\frac{q^{k-1}-1}{q^{2n}-1}.$ Taking a union bound over all nonzero vectors in $B_q^S(2n,d-1)$, we obtain
\[
\Pr\bigl[
\cC\cap B_q^S(2n,d-1)\neq\{\mathbf{0}\}
\bigr]
\le
\bigl(\Vol_q^S(2n,d-1)-1\bigr)
\frac{q^{k-1}-1}{q^{2n}-1}
\leq q^{2nH_{q^2}(\delta)+k-2n-1}\leq q^{n-2n+k-1}\leq \frac{1}{q^2}.
\]
So far we prove that with probability at least $1-\frac{1}{4}-\frac{2}{q^2}-e^{-\Omega(n)}\geq \frac{1}{4}-e^{-\Omega(n)}$, there exists $(k-1)$-dimensional symplectic orthogonal code $\cC$ such that $Z\leq L$, $\cC\cap B_q^S(2n,d-1)=\{\mathbf{0}\}$ and $\mathcal B(\cC)$ does not contain any $\cC$-dependent subset of size at most $L$. 

 Note that $Z\leq L$ forces that  $|\mathcal B(\cC)|\leq L$. Since we have already shown that $\mathcal B(\cC)$ does not contain any $\cC$-dependent subset of size at most $L$, this implies that $\mathcal B(\cC)$ is $\cC$-independent.
Lemma~\ref{lem:one-step-elimination} now yields a $k$-dimensional symplectic self-orthogonal code $\cC'\supseteq \cC$ such that $\cC'^{\perp_S}\cap B_q^S(2n,d-1)=\{\mathbf{0}\}.$ Hence, $d_S(\cC'^{\perp_S})\ge d$ and the proof is completed.
\end{proof}

Combining Lemma~\ref{lem:construct-quantum-codes} and Theorem~\ref{thm:quantum-gv-linear-factor}, we prove the existence of quantum codes beyond GV bound.

\begin{cor}
\label{cor:quantum-gv}
Fix a prime power $q$ and a constant $\delta\in(0,1-1/q^2)$ satisfying $H_{q^2}(\delta)<1/2$. Let $d-1=\lfloor\delta n\rfloor$. Then there exists a constant $c_{q,\delta}>0$ such that, for all sufficiently large $n$, a $q$-ary quantum code $[[n,n-k,d']]_q$ with $d'\ge d$ exists whenever
\[
\frac{q^{2n-k}-1}{q-1}<c_{q,\delta}n\frac{q^{2n}}{\Vol_q^S(2n,d-1)}.
\]
\end{cor}

\bibliographystyle{IEEEtran}
\bibliography{refs}  


\end{document}